\documentclass{emulateapj}
\usepackage[usenames,dvipsnames]{xcolor}
\definecolor{blue}{rgb}{0,0,1}
\usepackage{amsmath,amssymb,amsthm}
\usepackage{natbib}
\usepackage{graphicx}

\usepackage[hyperfootnotes=false]{hyperref}

\hypersetup{
  colorlinks,
  citecolor=blue,
  linkcolor=blue,
  urlcolor=Green}

\shorttitle{Resonances of Multiple Exoplanets and Implications for Their Formation}
\shortauthors{Zhang et al. 2014}

\begin{document}

\title{Resonances of Multiple Exoplanets and Implications for Their Formation}
\submitted{Accepted for ApJ Letters \ \ 2014 May 29}

\author{Xiaojia Zhang\altaffilmark{1}$^*$, Hui Li\altaffilmark{2}, Shengtai Li\altaffilmark{2}, 
Douglas N. C. Lin\altaffilmark{1,3}}
\altaffiltext{*}{E-mail: \href{mailto:xzhang47@ucsc.edu}{xzhang47@ucsc.edu}}
\altaffiltext{1}{Department of Astronomy and Astrophysics, University of California, Santa Cruz, CA 95064, USA}
\altaffiltext{2}{Los Alamos National Laboratory, Los Alamos, NM 87545, USA}
\altaffiltext{3}{Institute for Advanced Studies, Tsinghua University, Beijing, China}

\begin{abstract}
Among $\sim 160$ of the multiple exoplanetary systems confirmed, about $30\%$ of them
have neighboring pairs with a period ratio $\leq 2$. A significant fraction of these
pairs are around mean motion resonance (MMR), more interestingly,  peak 
around 2:1 and 3:2, with a clear absence of more closely packed 
MMRs with period ratios less than 4:3, regardless of planet masses. 
Here we report numerical simulations demonstrating that such 
MMR behavior places important constraints on the 
disk evolution stage out of which the 
observed planets formed. Multiple massive planets (with mass $\geq 0.8$ 
$M_{\rm Jup}$) tend to end up with a 2:1 MMR mostly independent of
the disk masses but low-mass planets (with 
mass $\leq 30$ $M_{\oplus}$) can have MMRs larger than 4:3 
only when the disk mass is quite small, suggesting that the observed 
dynamical architecture of most low-mass-planet pairs was established
late in the disk evolution stage, just before it was dispersed completely. 
\end{abstract}


\section{Introduction}

Many multiple planetary candidates were discovered by Kepler's 
transit search. It has been pervasively suggested that the 
majority of them are indeed genuine multiple-planet systems
\citep{2012ApJ...750..112L,2013ApJS..204...24B}. A significant fraction of adjacent 
pairs of planets are in or near 
mean motion resonances (MMRs) with period ratios around 2:1, 
3:2, 5:3, etc. \citep{2011ApJS..197....8L}, although 
members of most multiple systems do not have nearly commensurable 
orbits \citep{2009A&A...493..639M}. 

The MMR of exoplanet systems has been well studied under various 
perspectives \citep{1965MNRAS.130..159G,1980AJ.....85.1122W,1982CeMec..27....3H,
1985Icar...62...16W,2009DPS....41.4005O,2010ApJ...721.1184O,
2013ApJ...774...52L}. A widely adopted scenario is that resonant pairs 
captured each other on their mutual MMRs through convergent migration 
\citep{2000ApJ...540.1091B,2000MNRAS.313L..47K,2002astro.ph..9176L}. 
Migration mechanisms include tidal interaction between very short period 
planets and their host stars \citep{2010ApJ...724L..53S} as well as 
protoplanet interaction with their natal disks \citep{1996Natur.380..606L}.
For planets with periods longer than a few days, MMRs provide  
supporting evidence of the disk migration scenario.

Gas-giant planets with sufficient mass to open gaps in their natal disks
undergo Type II migration 
\citep{1986ApJ...309..846L}.  Most {\it Kepler} candidates have 
much lower masses and undergo Type I migration \citep{1980ApJ...241..425G,
1997ApJ...482L.211W,2011MNRAS.410..293P,2012ApJ...755...74K}. In 
either case, orbital convergence would occur if the inward migration
of a planet catches up with that of its siblings closer to the host star
or if a planet is trapped at some special disk region and its siblings
migrate toward it. Convergent migration leads to the possibility of
resonant capture \citep{2002ApJ...567..596L,2004A&A...414..735K}.  

Many multiple planet systems are indeed locked in or are close to MMRs. 
The current database on {\bf The Extrasolar 
Planets Encyclopaedia} Website (http://exoplanet.eu) contains $157$ confirmed 
multiple planet systems with $\sim 225$ neighboring pairs. This sample 
includes 118 systems of 2 planets, 22 systems of 3 planets, 9 systems 
of 4 planets, 4 systems of 5 planets and 4 systems of 6 planets.
Figure 1 shows the distribution of these adjacent pairs as a function 
of their period ratios, focusing on systems with a period ratio 
around and less than $4$. Overall, $\sim 70$ pairs (i.e., $\sim 1/3$ 
of all pairs) have a period ratio around or less than $2$. The subsets 
of pairs with {\it both} planets' masses $> 0.8 M_{\rm Jupiter}$ and 
below $\sim 30 M_{\oplus}$ (designated as gas-giant and low-mass-planet 
systems respectively) are also shown in Figure 1. Most gas giants 
are discovered by the radial velocity method.  Although their mass 
determinations are lower limits, this observational bias does not affect 
their statistics. Most multiple low-mass-planet systems are {\it Kepler} 
candidates. Their mass has been estimated by using an empirical formula 
to convert a planet's radius to mass
 \citep{2011ApJS..197....8L,2011Natur.470...53L}.  

\begin{figure}
\centering
\includegraphics[width=0.95\linewidth,clip=true]{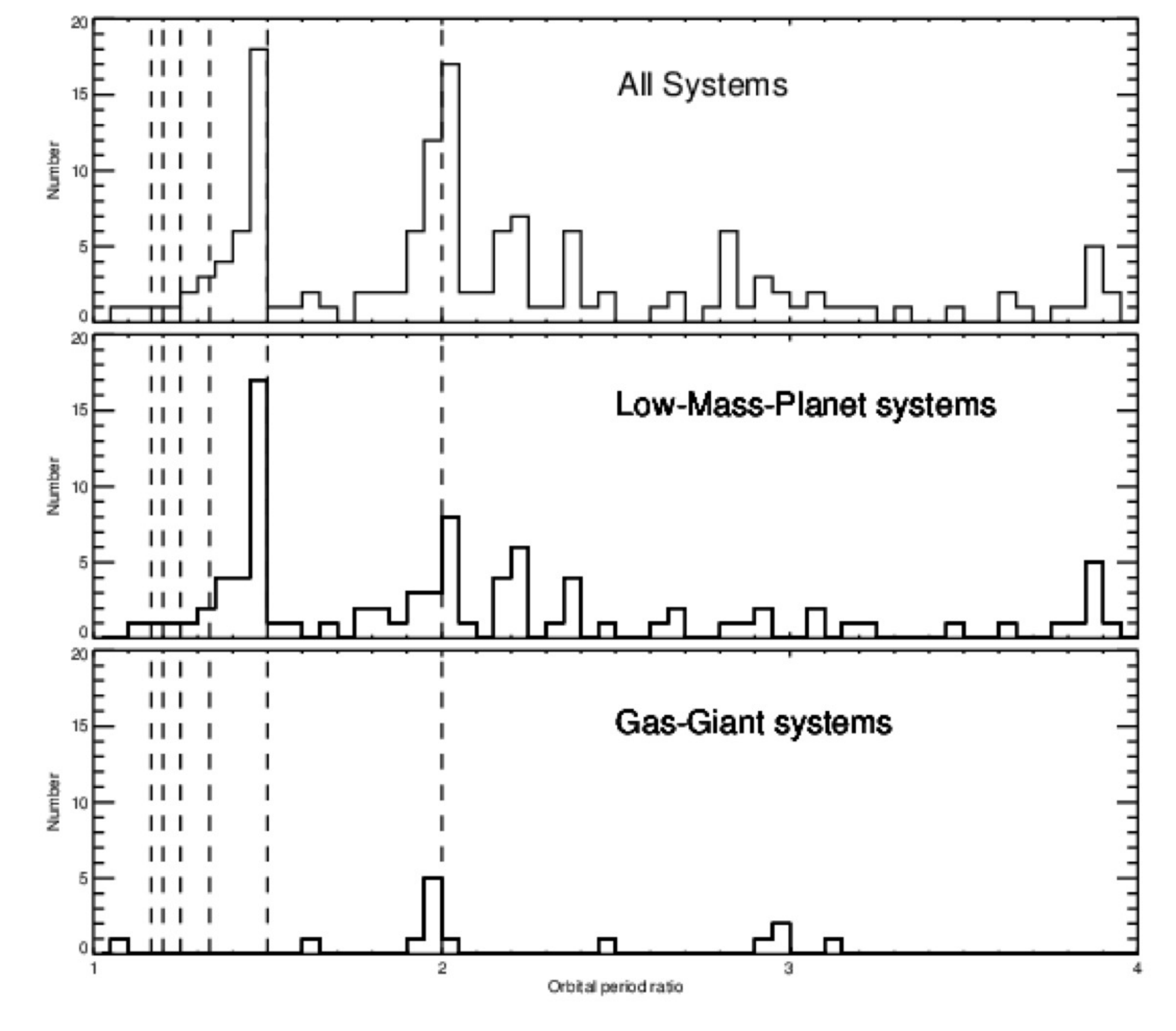}
\caption{
{\bf Top-bottom}: 
distribution of period ratios of adjacent pairs from 
all the confirmed multiple-planet systems {\bf (top)}, 
just the low-mass-planet systems with both planet 
masses $\leq 30 M_{\oplus}$ {\bf (middle)}, and 
just the gas-giant systems with both planet masses
$\geq 0.8 M_{\rm Jup}$ {\bf (bottom)}. 
The vertical dashed lines indicate the first-order MMR at 2:1, 3:2, 4:3, 5:4, 6:5 and 7:6, 
from right to left respectively.  Note that 
only pairs with a period ratio $\leq 4$ are shown, which
is about $70\%$ of all planet pairs in the 157 confirmed systems. 
Whereas most of the gas-giant systems do not have
MMRs smaller than 2:1, low-mass-planet systems can 
have MMRs down to 4:3, but there is a clear deficit 
of systems with a period ratio less than 4:3.}
\end{figure}

Figure 1 confirms the previously known results that there are statistically 
clusters of systems in or near the 2:1 and 3:2 MMRs 
\citep{2011ApJS..197....8L,2012DDA....43.0103F}.  Since the confirmed 
Kepler multi-exoplanetary systems measured with TTVs are biasing the samples toward
pairs near resonances, the peaks may be overrepresented, but one can still find
this statistical clustering feature by including all pairs of Kepler candidates
\citep{2014AJ....147...32G}.
The excess near the  
3:2 MMR is mostly associated with low-mass-planet systems, whereas that near the 
2:1 MMR contains both low-mass-planet and gas-giant systems. If these systems 
formed through resonant capture during planetary migration, it appears that 
their asymptotic MMR is a sensitive function of planet mass, orbital 
characteristics, and relative migration rates \citep{2013ApJ...775...34O}. 
Gas-giant pairs such as \objectname{GJ876}
\citep{2001AAS...199.0308L,2001ApJ...558..392R,2005ApJ...622.1182L}
and \objectname{HD82943} \citep{2006ApJ...641.1178L,2013ApJ...777..101T}
cluster mostly around the 2:1 MMR because the Type II migration rate
is relatively slow compared with the libration timescale for the 2:1 MMR.

However, Type I migration is considerably faster than Type II migration
\citep{1986ApJ...309..846L}. This allows migrating pairs to bypass wider resonances, with slower libration timescales
\citep{1999ssd..book.....M}, and end up in more tightly packed configurations. 
It is noteworthy that there is a deficit of both low-mass-planet
and gas-giant pairs with period ratios smaller than 4:3.  Most {\it Kepler} 
planetary candidates have orbital periods within a few months, and the
possibility of not detecting a transiting companion with similar periods 
is small. Thus, the observed deficit of closely packed resonant 
pairs and the preferential concentration of lower-mass pairs near the 3:2
MMR relative to the 2:1 MMR are statistically significant.

\section{Method}

In order to reproduce these observations in terms of the 
resonant-capture scenario, we use a two-dimensional (2D) module
of the LA-COMASS package developed at Los Alamos 
to investigate the orbital evolution of multiple planets in 
a protostellar disk. This 2D hydrodynamical 
polar grid code solves the continuity and isothermal Navier-Stokes 
equations for gas inside a quasi-Keplerian disk subject 
to the gravity of a one-solar-mass central star. The 
motion of multiple planets is calculated with a fourth-order 
Runge-Kutta solver. The evanescent boundary condition 
has been implemented in the code to provide wave killing zones 
at each edge of the disk \citep{2006MNRAS.370..529D}.
This code has been used extensively for planet-disk interaction studies
(e.g., \cite{2009ApJ...690L..52L}).

In our simulations, the 2D disk is modeled within the radial 
range of $[0.3R_0,3.3R_0]$ where $R_0$ is the distance unit in the code. 
We investigate two main configurations: 
a disk with multiple gas giants and a disk with multiple 
$10 M_{\oplus}$ planets. 
For the migration of gas giants, 
since their Type II migration does not 
sensitively depend on the disk structure \citep{1986ApJ...309..846L}
, we adopt models
with smooth density, temperature and viscosity profiles. However,
the pace and direction of the Type I migration 
does depend sensitively on the disk properties and they may be 
trapped in a disk near the magnetospheric truncation radius, the 
inner edge of "dead zone" or other locations such as the 
boundary between the viscous heating and irradiation heating 
region \citep{2012ApJ...755...74K}.

For the migration of low-mass planets, 
we adopt a two-zone disk model based on the $\alpha$ prescription
for viscosity.  In order to approximate a disk structure in which
the magnetohydrodynamic turbulence is prevalent throughout the inner region and
the mostly neutral outer regions with a ``dead'' midplane zone,
we adopt a high value of $\alpha_{vis}=0.004$ for disk radius 
$a < a_{crit}=0.6R_0$ and $\alpha_{vis}=0.001$ for $a > a_{crit}$.
Typical values of $R_0$ may range from a few stellar radii (for
the inner disk boundary or the inner boundary of the dead zone) 
to a few AUs (for the interface between the viscously heated and 
irradiated disk regions).  In a steady state \citep{2009ApJ...690..407K}, 
this prescription leads to a disk surface density ($\Sigma$) and 
pressure enhancement across $a_{crit}$.  This $\Sigma$ distribution 
affects the direction and pace of low-mass-planets Type I 
migration \citep{2010MNRAS.401.1950P,2011MNRAS.410..293P}. 
The planets are placed relatively far away from $R_0$ so that their initial migration
 follows the typical Type I migration.

The ratio of disk scale height over disk radius is taken to be a 
constant of $h/r = 0.05$.  For all simulations, the initial semi-axis of planet pairs are 
all beyond $R_0$. 
The planets are held on their initial 
orbits for 1000 Keplerian periods and are released 
after their disk has adjusted to their perturbation. Our simulated 
time span is much shorter than the Myr disk depletion timescale. The overall 
change in the disk mass is modest.

\section{Simulation Results}

The top panel of Figure 2 shows the evolution of 
the period ratio of two gas giants under a combination of two 
different mass ratios and two different sets of initial radial 
separations, based on the assumption that gas giants are 
formed independently and separated by modest period ratios 
($\sim 3$). Since the density waves 
excited by the two gas giants propagate throughout the remaining 
ring between them, gas diffuses through the gaps around each 
planet, in contrast to the shepherded planetary rings.  After gas 
in the ring is depleted, the exchange of angular momentum 
between planets and inner/outer disks induces them to converge
\citep{2000ApJ...540.1091B}.  After a few thousand orbits, these 
gas giants capture each other into a 2:1 MMR.  For the gas-giants pairs,
we only simulated the convergent migration using a typical set of disk conditions.
Since the Type II migration timescale is comparable to or even longer than the viscous 
evolution timescale of the disk, the gas giants can hardly pass through the 2:1 MMR 
location unless they are in a disk with extremely high accretion rate.

\begin{figure}
\centering
\includegraphics[width=0.85\linewidth,clip=true]{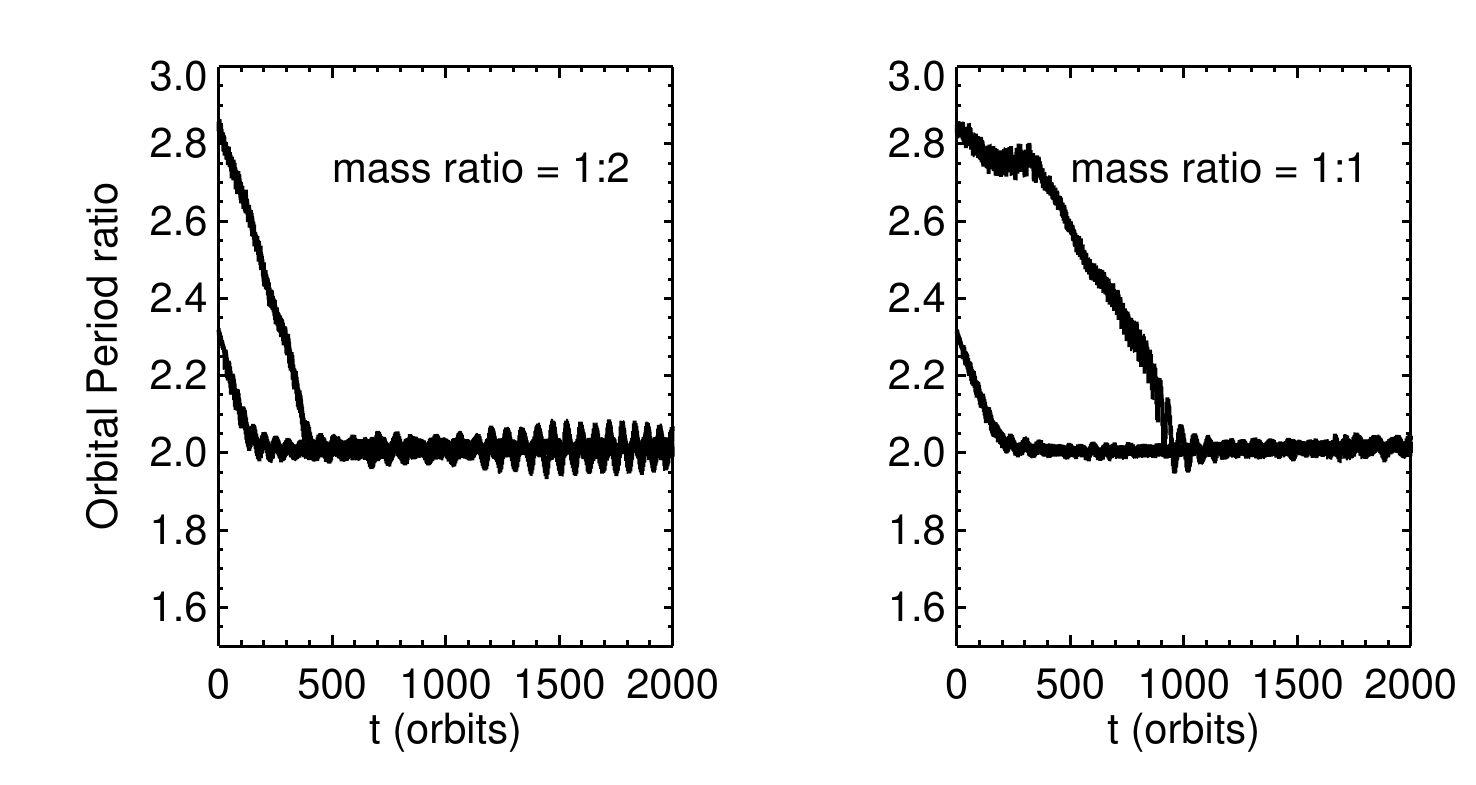}
\includegraphics[width=0.95\linewidth,clip=true]{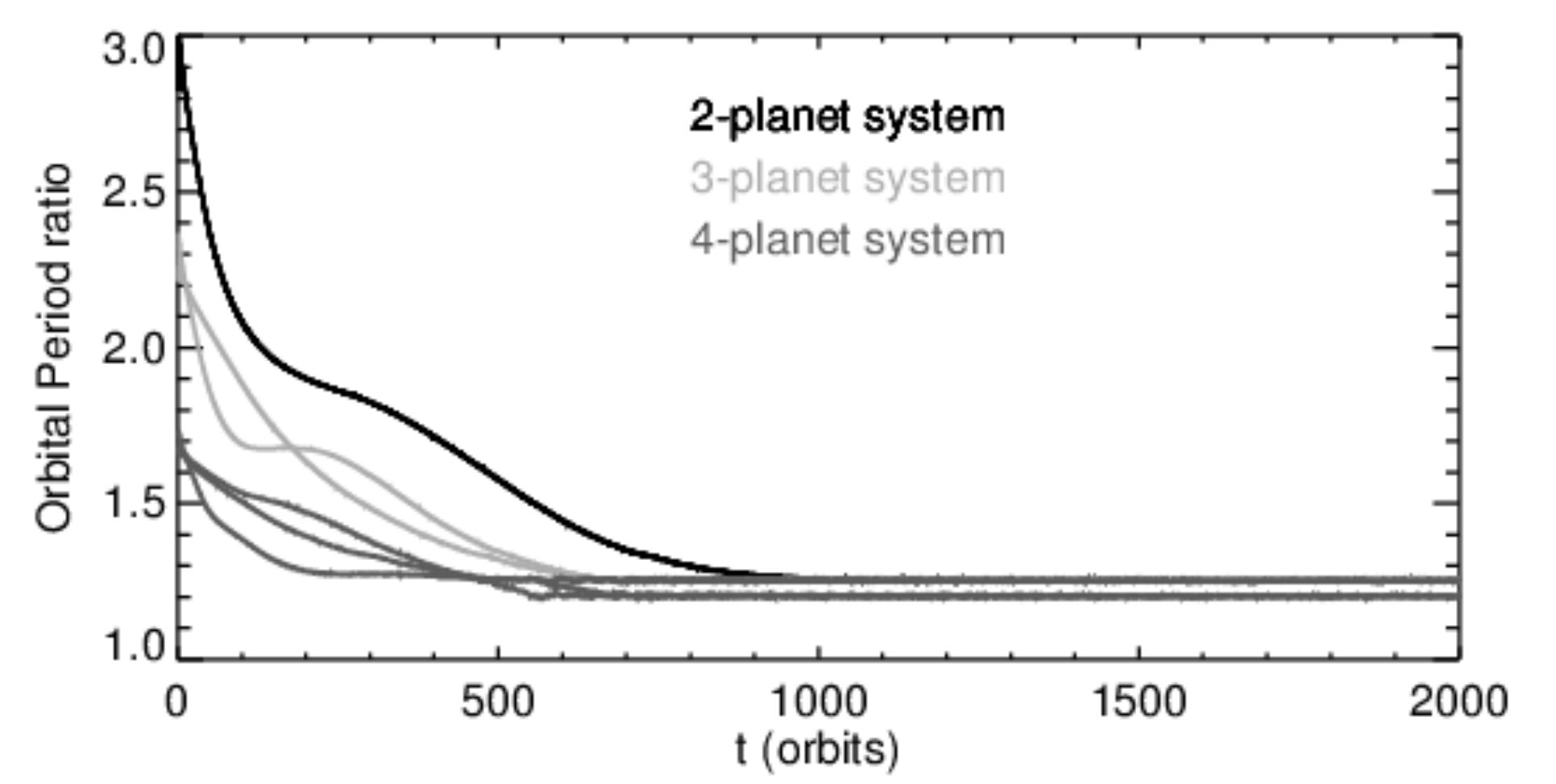}
\caption{
{\bf Top}: the 
curves represent the period ratio evolution of gas-giant pairs 
with combinations of outer-to-inner planet mass ratios of $0.5$ (left panel)
and $1$ (right panel), and initial period ratio of 2.83 and 2.31. All planets are 
released initially with circular orbits. The inner planet mass is fixed 
at $2M_{J}$. They all settle to a period ratio of around 2:1. 
{\bf Bottom}: the period ratio evolution of low-mass-planet pairs in 
two-planet (black curve), three-planet (light gray curves) and four-planet (dark gray curves) 
systems. All planets have $10 M_{\oplus}$. They all settle to a period ratio of 
around 5:4 or 6:5.}
\end{figure}

We also simulated the evolution of initially compact gas-giant pairs
(with a period ratio less than 2). Population synthesis models 
\citep{2013ApJ...775...42I} indicate that multiple gas-giant 
systems can form with relatively compact orbits because their 
progenitor embryos are separated by $\sim 10$ Hill's 
radii before they acquire sufficient mass to accrete gas 
efficiently.  Our extensive simulations 
show that the orbits of these compact gas-giant pairs are not stable 
against their intense gravitational perturbation on each other
\citep{2007ApJ...666..423Z}.  In all cases, one of the two gas giants 
is scattered to large distances from its initial location.  In the 
inner regions, the scattered planets may resume their migration.
If they are able to catch up with their companions, they would eventually 
settle into the 2:1 resonance.  The 
gas-giants' resonance ``barrier'' at the 2:1 MMR is confirmed by these 
simulations.

Typically the Type I migration timescale of a planet on a circular orbit is 
$\tau_0 \sim h^2 q^{-1} M_{\star}/(\Sigma_p R_p^2) \Omega_p^{-1}$, where 
$h$ is the aspect ratio of the disk, $q$ is the mass ratio of planet to star, 
$R_p$ is the location of planets and 
$\Sigma_p$ is the local surface density at $R_p$. The migration rate also sensitively dependent on
the gradient of the surface density and temperature of the disk \citep{2010MNRAS.401.1950P,2011MNRAS.410..293P}. In this Letter, we only control the 
migration rate by applying different surface densities.
Our disk model leads 
to the convergent evolution between adjacent pairs of two, three, or four 
low-mass-planet systems.  In these
simulations, the mass of each planet is assumed to be $10 M_\odot$ 
and the gas mass within the computational domain is $2 \times 10^{-3}$ 
that of the host star.  The asymptotic period ratio of all neighboring 
pairs is around 5:4 or smaller. This ability to closely pack multiple 
low-mass-planet systems is due to their relatively rapid converging speed.

Such compact systems are rarely found among the Kepler planetary 
candidates.  This apparent discrepancy between observations and 
simulations can be reduced considerably with much slower converging
Type I migration rates.  In order to illustrate this conjecture, we have performed 
a large set of simulations of multiple low-mass systems for a broad 
range of disk surface densities. 

In these simulations, we have chosen 
the central star mass  $M_{\star}= 1 M_{\odot}$, $h/r=0.05$ and 
$\alpha_{vis}=0.004$/0.001 for $a$ interior/exterior to $a_{crit}$. 
In a steady state, the disk accretion rate is given by
$\dot{M}=1.5\times 10^{-4} \cdot f \cdot (\frac{1AU}{R_0})^{1.5} 
M_{\odot}/yr$, where $f=\frac{\Sigma_0 R_0 ^2}{M_{\star}}$ is the 
disk-to-star mass ratio and $\Sigma_0$ is the disk surface density 
(in $g/cm^2$) at $R_0$. 
The transition radius $a_{crit}=0.6R_0$ is also dependent on the disk 
accretion rate \citep{2009ApJ...690..407K} as $a_{crit} \propto 
\dot{M}^{0.5}$.  Both the disk surface density normalization and 
the trapping location (associated with $R_0$) decrease as the disk 
accretion rate diminishes during disk depletion.  For example, 
with $f = 0.002$, we have $\dot{M}=3.6\times 10^{-8} M_{\odot}/yr$ 
and $a_{crit}=2.5AU$.  
 In our simulations, there is no time evolution 
of the disk, except for one case shown in Figure 4. In other cases, 
the accretion rate is constant along the simulations. 
Varying $\dot M$ is equivalent to varying disk mass which is equivalent to 
varying the establishing time of the planet orbital architecture.

From these simulations, we find that
the asymptotic period ratios between adjacent pairs decrease toward unity 
as the accretion rate increases. Figure 3 shows a general trend that lower disk 
accretion rates lead to relatively wider spacing for low-mass-planet systems.
These results suggest that the accretion rate must be $< 2 \cdot 
10^{-8} M_{\odot}/yr$ to reproduce the observed paucity. If this paucity is a signature of some alternative dynamical process, like instability,  then we can constrain the forming stage of the planetary systems from the accretion rate corresponding to the final period ratio larger than the 4:3 MMR. The corresponding disk-to-star mass ratio in 
this region is $< 6.7\times 10^{-4}$.  The reproduction
of the observed enhancement of adjacent low-mass-planet pairs with 3:2 MMR 
requires even lower disk accretion rates ($\dot{M} < 10^{-8} M_{\odot}/yr$) 
especially in systems with more than two planets. Note that the inferred 
accretion rates around classical T Tauri stars in the Taurus and Ophiuchus
complex \citep{2006A&A...452..245N} range mostly between 
$10^{-8}-10^{-7} M_{\odot} yr^{-1}$.

\begin{figure}
\centering
\includegraphics[width=0.95\linewidth,clip=true]{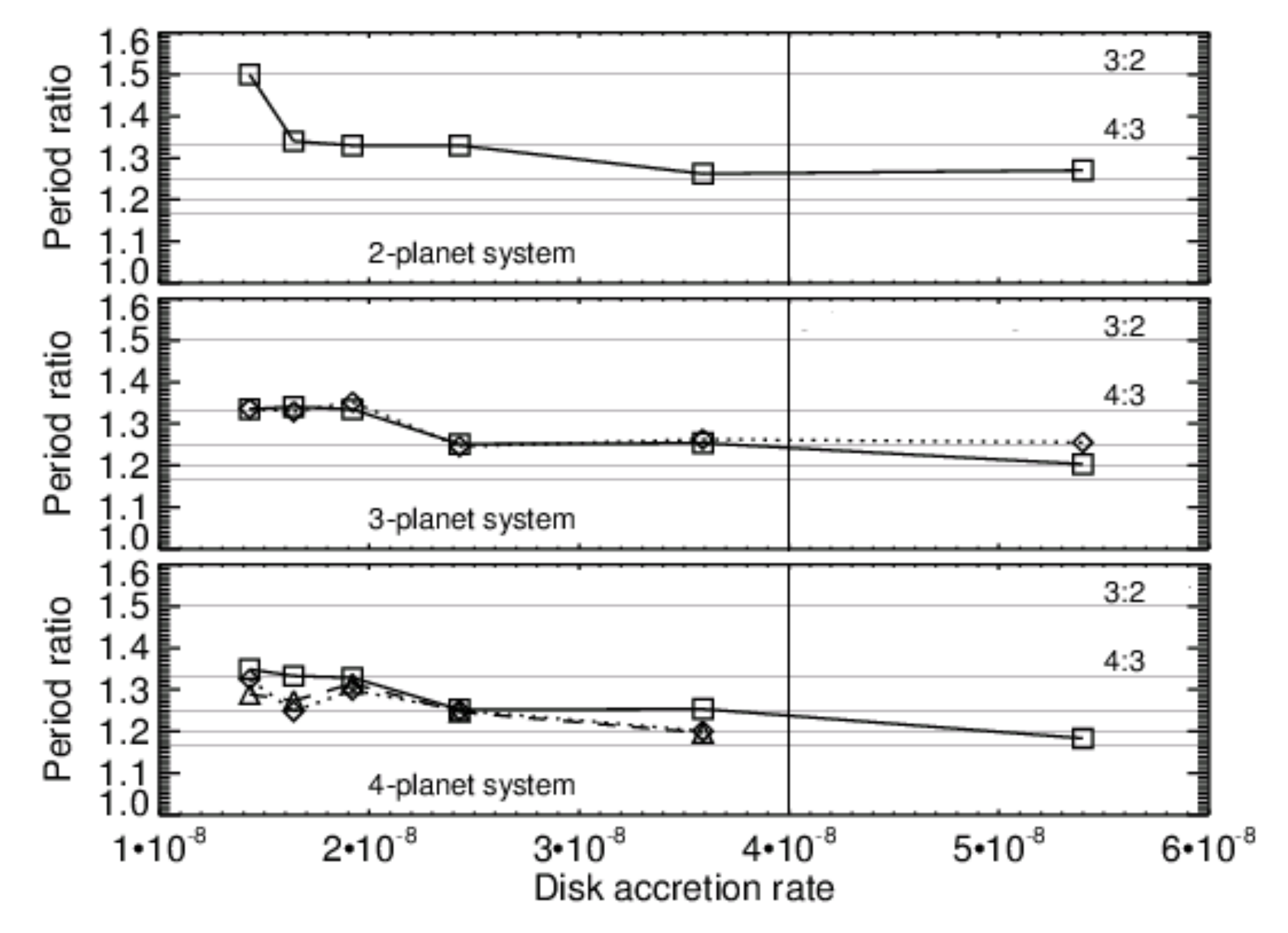}
\caption{
Panels from top to bottom show the final period ratio of adjacent 
pairs in multiple low-mass-planet systems with two, three and four planets 
of $10 M_{\oplus}$ each. The horizontal lines indicate the 
first-order MMRs at 3:2, 4:3, 5:4, 6:5 and 7:6. 
The vertical line marks the observed disk accretion rate 
for a typical young solar mass system with an age 
$\sim 1$ Myr \citep{2006A&A...452..245N}.}
\end{figure}

In order to extract constraints on the disk accretion rate from Figure 1, 
we need to verify that the stability of compact systems may be preserved
\citep{1993Icar..106..247G}. 
Systematic studies  \citep{2007ApJ...666..423Z} show that in a gas-free 
environment, multiple equal-mass low-mass-planet systems with initially 
circular orbits separated by $k_0 = \Delta a / R_H < 10$ (where $R_H = 
(2 M_p/3M_\ast)^{1/3} a$) become dynamically unstable over the Gyr 
main sequence lifespan of the host stars.  For non-resonant 
$M_p=10 M_\oplus$ low-mass-planet systems, this stability criterion 
requires the period ratio to be larger than $\sim 1.43$, which could 
account for the paucity of multiple systems with a period 
ratio between adjacent planets smaller than that of the 3:2 MMR. 
Multiple gas-giant systems with a period ratio between 
adjacent planets smaller than that of the 2:1 MMR are also unstable 
on Gyr timescale. 

However, the stability of systems with near MMRs may be better preserved against 
long-term dynamical instability. 
To illustrate, we select a pair 
of low-mass planets that are captured into each other's 5:4 MMR in a disk 
initially with $f=0.002$.  Such a system corresponds to $k_0=6$.  If it
is out of resonance, it would become dynamically unstable in $<10^3$ 
orbits in the absence of gas and $<10^4$ orbits if it is embedded in 
a minimum mass solar nebula.  We extend our simulation for an additional
$10^4$ orbits while the disk surface density is prescribed to decrease 
exponentially over that timescale.  The results in Figure 4 indicate 
that once the low-mass-planet pairs are captured into a tight MMR, 
they tend to remain in these MMRs in the absence of major perturbations before the 
depletion of the disk. 
However, the later long-term orbital evolution of planets with an absence of gas 
may break the system. Although we can not exclude the instability criterion as an contribution to the
paucity of a period ratio smaller than the 4:3 MMR, we can take this as a clue to the formation stage. If instability is responsible for the paucity, it might indicate that
 the low-mass-planet pairs with a period ratio larger than 4:3 formed in a somewhat late stage of the disk.

\begin{figure}
\centering
\includegraphics[width=0.99\linewidth,clip=true]{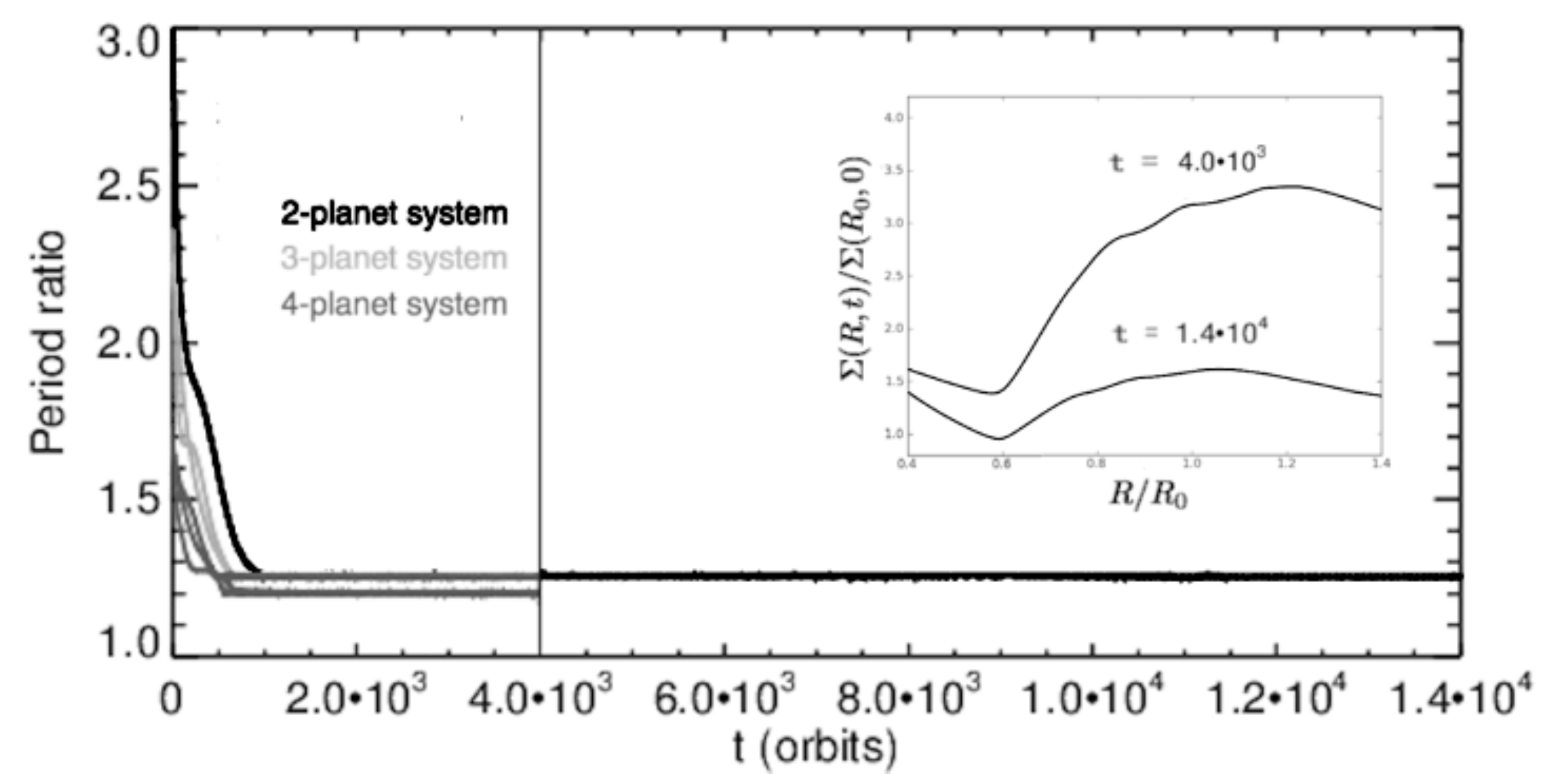}
\caption{
Pairs of low-mass-planets undergo Type I migration and capture each other 
within $10^3$ orbits, as indicated by the left panel. Continued 
simulations up to $10^4$ orbits of such systems demonstrate that 
the pair remains around the 5:4 MMR. 
A gradual decrease of the disk surface density is implemented, 
as shown in the insert.}
\end{figure}

Based on the results from Figures 3 and 4, we infer that, low-mass-planet 
systems can capture each other into 4:3 and/or 3:2 MMR,s provided they
undergo Type I migration when the accretion rate 
($\dot M$) or the mass of their natal disks is relatively low.  
The results in Figure 3 indicate that in order for a pair of $10 M_\oplus$ 
low-mass planets to attain an asymptotic period ratios of 4:3 and 3:2,
the upper limits of the disk-to-star mass ratios are at 
$f=6.7\times 10^{-4}$ and $f=4 \times 10^{-4}$, respectively. 
Since the Type I migration rate is linearly proportional to the product of
$q$ and $f$, low-mass-planet pairs would approach each other's 4:3 
(or 3:2) MMRs with the same critical relative speed if their 
$q\cdot f \sim 2\times 10^{-8}$ (or $\sim 1.2\times 10^{-8}$).  
The condition for MMR capture requires that the migration timescale through 
the characteristic width ($\tau_{mig} \sim \Delta a / \dot a$) is longer than the 
libration timescale ($\tau_{lib}$).
If we take into account that the libration time  of the lowest-order 
MMR is $\propto q^{-1/2}$ and $\Delta a \propto q^{1/2}$ in the first order of the expansion
\citep{1999ssd..book.....M}, the critical condition for the lowest-order of 
MMR capture would be roughly independent on the planet-to-star mass ratio.

In order to quantify the stage of the disk according to the accretion rate, we use the observations of a 
relatively young ($t_a \sim 1$ Myr) star forming region, $\rho$ 
Oph \citep{2006A&A...452..245N}, where the accretion rate around 
$\sim 1 M_\odot$ star is about $4\times 10^{-8}M_{\odot}/yr$.
For $R_0=4.1AU$, the corresponding value $f=0.002$. 
The accretion rate is also observed to decline over a timescale of
3-5 Myr.  We introduce an approximation for the disk accretion rate
$\dot{M}=4\times 10^{-8} \cdot e^{-(\frac{t-t_a}{\tau_D})}M_{\odot}/yr$
for $t \geq t_a$, 
where $t$ indicates the age of the disk, $t_a \sim 1~Myr$, and $\tau_D$ 
is the disk lifetime. Figure 5 shows the dependency of final period 
ratios on the planet masses and the MMRs formation times. 
It confirms that a relatively large period ratio (such as 3:2)
requires the low-mass-planet pairs to have migrated and captured each other
late in the disk evolution stage when the disk surface 
density was sufficiently depleted.

\begin{figure}
\centering
\includegraphics[width=0.95\linewidth,clip=true]{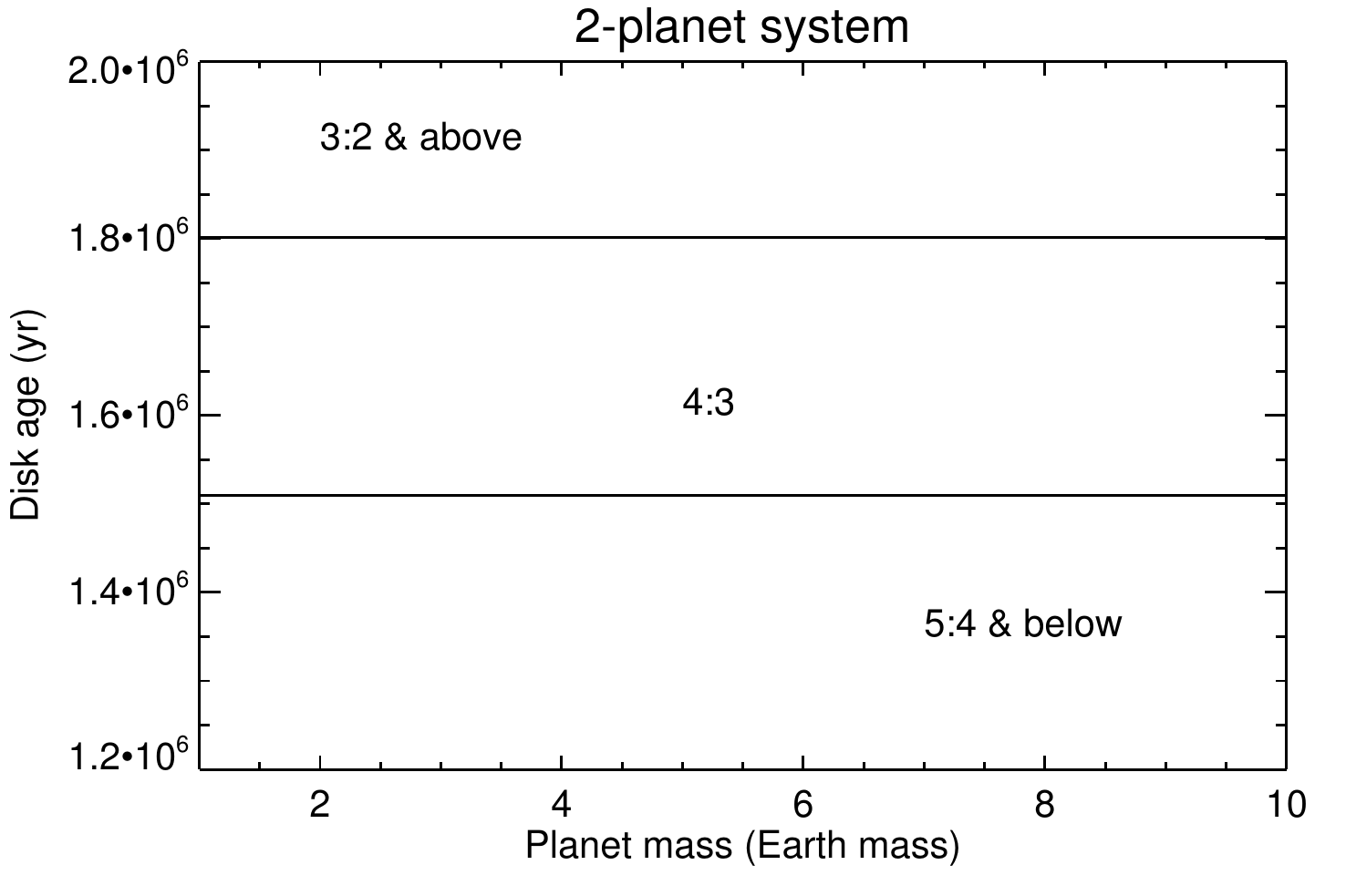}
\caption{
Two solid lines, from top to bottom, mark the earliest formation time for
planet pairs that produce final period ratios of 3:2 and 4:3, respectively. 
An exponentially decreasing disk accretion rate model is used in this calculation.}
\end{figure}

\section{Summary}

These simulation results for multiple gas-giant and low-mass-planet 
systems place important constraints on the planet formation stage with 
respect to the disk evolution. According to the core accretion 
scenario, the formation of gas giants must be preceded by the emergence
of sufficiently massive ($>10 M_\oplus$) protostellar embryos in a gas-rich environment, presumably during the early stage of disk evolution.
These cores, if retained in a dense disk, would either congregate and 
effectively merge near some trapping radius to become cores of 
proto-gas-giant planets or be mostly scattered into or far from their 
host stars.  A significant ``left-over'' population would have produced 
compact pairs with small period ratios that are not consistent with 
observations. Around stars that only bear relatively low-mass 
planets, their dynamical configuration may be established 
during the advanced stages of disk evolution when the disk gas
is severely depleted.  It is also possible that these low-mass planets
were assembled over several millions of years.

\acknowledgments

We thank the referee, Frederic Rasio, for helpful comments that improved the manuscript.
We acknowledge support from the LDRD program and IGPP
 of Los Alamos National Laboratory. H.L. and D.N.C.L. also 
acknowledge support from the UC-fee program of University of California. 
Simulations were carried out using the Institutional Computing resources
at LANL.

\bibliographystyle{apj}
\bibliography{sample}

\end{document}